\documentstyle[aps,twocolumn]{revtex}     
\begin{document}

\newcommand{\be}{\begin{equation}}
\newcommand{\ee}{\end{equation}}
\newcommand{\bea}{\begin{eqnarray}}
\newcommand{\eea}{\end{eqnarray}}
\newcommand{\nn}{\nonumber}

\draft
\preprint{
\begin{minipage}{5cm}
MPG-VT-UR ../98\\
E2-98-...
\end{minipage}}

\title{On the Groundstate of Yang-Mills Quantum Mechanics}
\author{A M. Khvedelidze \ $^a$
\thanks{Permanent address: Tbilisi Mathematical Institute,
380093, Tbilisi, Georgia}
\, and \, H.-P. Pavel\ $^b$}
\address{$^a$ Bogoliubov Theoretical Laboratory, Joint Institute for Nuclear
Research, Dubna, Russia}
\address{$^b$ Fachbereich Physik der Universit\"at Rostock,
              D-18051 Rostock, Germany}
\date{\today}
\maketitle

\begin{abstract}

A systematic method to calculate the low energy spectrum
of $SU(2)$ Yang-Mills quantum mechanics with high precision is given
and applied to obtain the energies of the groundstate and the first few 
excited states.
\end{abstract}

\bigskip

\bigskip


\bigskip
\bigskip

The so-called Yang-Mills mechanics originates from Yang-Mills field theory
under the supposition  of the spatial homogeneity of the gauge  fields. 
Since its  introduction  twenty years ago  \cite{BasMatSav} 
this model has been studied  extensively from different points of view 
\cite{AsatSav} - \cite{Govaerts}.
The model is known to exhibit stochastical
behaviour at the classical level.
In particular it has been shown for a simplified two-dimensional version 
of the model that it is nonintegrable 
\cite{Shur}. How this reflects itself on 
the corresponding quantum mechanical problem is still 
under investigation \cite{Simon} - \cite{Medvedev} , \cite{Halperin}, 
\cite{PauseHeinzl}.
Furthermore it offers the possibility to develop methods 
generalizable to the study of the corresponding non-Abelian gauge field 
theory.
With this point of view Yang-Mills quantum mechanics 
has been investigated in the small volume approximation \cite{Luescher},
in the small coupling limit on the lattice using the 
semiclassical approximation \cite{BartBrunRaabe}, and to  
estimate the glueball spectrum \cite{Giga}.
Furthermore the question of the existence of normalizable zero-energy 
wave functions in supersymmetric matrix models \cite{deWit}  as 
generalizations of Yang-Mills mechanics is under intensive investigation 
(see e.g. \cite{Froehlich} and references therein).
   
In the present work we would like to investigate the groundstate
and the first few excited states of
$SU(2)$ Yang-Mills quantum mechanics using the variational approach.
The quantum treatment of such models is confronted with the difficulties 
due to the existence of the non-Abelian Gauss law constraints and the  
presence of the unphysical degrees of freedom. 
In  \cite{GKMP} it has been shown that classical $SU(2)$ Yang-Mills mechanics,
after the elimination of all unphysical degrees of freedom,
reduces to the equivalent unconstrained system described by the Hamiltonian  
\be
\label{eq:PYM}
H = \frac{1}{2}\sum_{cyclic}^3 \left[ p^2_i 
+\xi^2_i \frac{x_j^2 +  x_k^2}{\left(x_j^2 -
x_k^2\right)^2} + g^2  x_j^2 x_k^2\right]\,.
\ee
This Hamiltonian is defined on the phase space  spanned by the three 
canonical pairs $x_i, p_i$ and the three $SO(3)$ left-invariant Killing 
vectors $\xi_i$ satisfying the Poisson bracket relations
$\{\xi_i,\xi_j\}=-\epsilon_{ijk}\xi_k$~.
Each of the three configuration space variables  $x_i$ is defined on the 
positive half line.
The Hamiltonian (\ref{eq:PYM}) has been obtained by writing the spatially 
homogeneous $SU(2)$ gauge fields $A^a_{i}(t)$ in the so-called 
polar representation \cite{Marcus}  
\bea
\label{eq:pcantr}
A_{ai} \left(q, Q\right)
= O_{ak}\left(q\right) Q_{ki}~,\nn
\eea
with the orthogonal $3\times 3$ matrix \(O\) parametrized by the three 
angles $q$ and the positive definite $3\times 3$ symmetric matrix \( Q \).
The three angles $q$ are pure gauge degress of freedom, 
which do not enter the  Hamiltonian while  their conjugated momenta 
are vanishing according to  the  Gauss law constraints.
The variables $x_i$ appearing in the Hamiltonian
(\ref{eq:PYM}) are the eigenvalues  of the positive definite 
symmetric matrix \( Q\)   
\bea
Q = R^{T}(\chi)D (x_1,x_2,x_3) R(\chi)~.\nn
\eea
The angles \(\chi\) in the orthogonal matrix $R(\chi)$ and their canonical 
momenta are combined in the Killing vectors $\xi_k$~. 
We emphasize that by definition of the polar decomposition 
the diagonal elements $x_i$ are positive definite which has
important consequences on the quantum level, as will be shown below.

The transition to the corresponding quantum Hamiltonian 
is free of operator ordering ambiguities.
The operators  $J_i=R_{ij}\xi_j$ 
corresponding to the right-invariant $SO(3)$ Killing vectors 
commute with the Hamilton operator. Its eigenstates can therefore
be classified according to the quantum numbers $J$ and $M$ as eigenvalues 
of the total spin  $\vec{J}^2$ and its projection $J_3$.
We shall consider here only the spin zero sector, $\vec{J}^2=\vec{\xi}^2=0$,
for which the Schr\"odinger eigenvalue problem reduces to 
\be
\label{H--0}
\left[-{1\over 2}\sum_{i=1}^3{\partial^2\over\partial x_i^2}
+{g^2\over 2}\sum_{i<j} x_i^2 x_j^2 \right]\Psi_E
= E \Psi_E~.
\ee
For the Hermiticity of the Hamiltonian $H_0$ defined in (\ref{H--0}) 
the wave functions have to satisfy the boundary conditions
\bea
\label{bc1}
&&\lim_{x_i\rightarrow 0}\Psi(x_1,x_2,x_3)=0~,\\
&&\lim_{x_i\rightarrow\infty}\Psi(x_1,x_2,x_3)=0\,.\label{bc2}
\eea
The potential in (\ref{H--0}) has three flat valleys 
of zero  energy,  $x_1=x_2=0\ $ and  $  x_3 \,   {\rm arbitrary}$,
and the two others obtained from this by cyclic permutation.
Close to the bottom of the  valleys the
potential is that of a harmonic oscillator with a 
width narrowing down for larger values of $x_3$.

One of the main results known for such a problem is that it has been proven
to have a discrete spectrum due to quantum fluctuations \cite{Simon}
\cite{Luescher}, 
although the classical problem allows for scattering trajectories.
Based on the  well-known operator inequality 
\footnote{The number 3 in r.h.s of the inequality originates from
the positiveness of $x$.}
\bea
-{\partial^2\over\partial x^2} + y^2x^2 \ge 3|y|~,\nn
\eea
it follows that
\be
\label{HH_0}
H_0\ge {1\over 4}\left(-\Delta + 3\sqrt{2}g(x_1+x_2+x_3)\right) 
=:{1\over 2} H^\prime~.   
\ee
Since the Hamiltonian $H^\prime$ is known \cite{Simon} 
to have a discret spectrum, this is true also for $H_0$.   
An important open question is at which energy the groundstate is.
The well-known nonnormalizable zero energy solution
$\Psi_{E=0}:=A\exp[-gx_1x_2x_3]$ \cite{Loos},\cite{KP}, does not
satisfy the necessary boundary condition (\ref{bc1}).
The knowledge of the groundstate energy of $H^\prime$ in inequality 
(\ref{HH_0}) would provide a lower bound for the groundstate energy of $H_0$.
Due to the additive structure of the potential term in $H^\prime$
one can make a separable ansatz for the solution of the corresponding
eigenvalue problem. The energy of the lowest such separable
$H^\prime$ eigenstate satisfying the above boundary conditions (\ref{bc1}) and 
(\ref{bc2}) is
\be 
\label{sepest}
E^{\prime}_{\rm sep} = 6|\xi_0|(3g/2)^{2/3}= 9.1924~ g^{2/3}~,
\ee
where $\xi_0 = -2.3381 $ is the first zero of the Airy function.
We shall see in this letter that there exist nonseparable solutions
which have an energy below (\ref{sepest}). 

To obtain an upper bound ${\cal E}$ for the groundstate energy 
$E_0$ of the Hamiltonian the  most  powerful  tool is the Rayleigh-Ritz 
variational technique  \cite{ReedSimon},
based on the minimization of energy functional 
\be
\label{energyf}
{\cal E}[\Psi] := \frac{\big<\Psi|H_0|\Psi\big>}
{\big<\Psi|\Psi\big>}~.
\ee 
Guided by the harmonic oscillator form of the valleys of the potential
in (\ref{H--0}) close to there bottom a simple first choice for a trial 
function compatible  with the boundary conditions (\ref{bc1}) and 
(\ref{bc2}) is to use the lowest state of three 
harmonic quantum oscilators on the positive half line      
\be
\label{Psi000}
\Psi_{000}=
8\prod_{i=1}^3 \left({\omega_i\over \pi}\right)^{1/4}\sqrt{\omega_i}x_i
e^{-\omega_ix_i^2/2}~.
\ee
The stationarity conditions for the energy functional 
of this state,   
\bea
{\cal E}[\Psi_{000}]=
\sum_{cyclic}^3 \left({3\over 4}\omega_i+
       {9\over 8}g^2{1\over \omega_j\omega_k}\right)~,\nn
\eea
lead to the isotropic optimal choice  
\be \label{fr}
\omega :=\omega_1=\omega_2=\omega_3=3^{1/3}g^{2/3}~.
\ee
As a first upper bound for the groundstate energy of the Hamiltonian we 
therefore find
\bea
%\label{1stest}
E_0 \le {\cal E}[\Psi_{000}]=
{27\over 8}3^{1/3}g^{2/3} = 4.8676~ g^{2/3}.\nn
\eea
This upper bound is in agreement with the lower bound of the energy
functional for separable functions   
\be 
\label{sepestPsi}
{\cal E}[\Psi_{\rm sep}] 
\ge {1\over 2} E^{\prime}_{\rm sep}= 4.5962~ g^{2/3}
\ee
obtained from the operator inequality (\ref{HH_0}) and the
lower bound (\ref{sepest}) for separable solutions of $H^\prime$.

In order to improve the upper bound for the groundstate energy
of the Hamiltonian $H_0$ we extend the space of trial functions 
(\ref{Psi000}) 
and consider the Fock space of the orthonormal set  of the products 
\be 
\label{bel}
\Psi_{n_1 n_2 n_3}:=
\prod_{i=1}^3 \Psi_{n_i}(\omega, x_i)~,
\ee
of the odd eigenfunctions of the harmonic oscillator
\bea
\Psi_{n}(\omega,x):=
            {(\omega/\pi)^{1/4}\over \sqrt{2^{2n}(2n+1)!}}
e^{-\omega x^2/2}
            H_{2n+1}(\sqrt{\omega}x)~,\nn
\eea
with the frequency fixed by (\ref{fr}).
Furthermore the variational procedure becomes much more effective,
if the space of trial functions is decomposed into the irreducible
representations of the residual discrete symmetries of the Hamiltonian 
(\ref{H--0}). 
It is invariant under arbitrary permutations of any two of the 
variables $P_{ij}x_i = x_jP_{ij}, \, P_{ij}p_i =p_jP_{ij}$ and 
under time reflections $Tx_i = x_iT,\,\,Tp_i = - p_iT$, 
\bea
%\label{TPsym}
[H_0,P_{ij}]=0,\,\,\,\,\qquad [H_0,T]=0\,.\nn
\eea
We shall represent these by the permutation operator $P_{12}$, 
the cyclic permutation operator $P_{123}$ and the time reflection
operator $T$, whose action on the states is 
\bea
P_{123}\Psi(x_1,x_2,x_3)&=&\Psi(x_2,x_3,x_1)~,\nn\\ 
P_{12}\Psi(x_1,x_2,x_3)&=&\Psi(x_2,x_1,x_3)~,\nn\\
T\Psi(x_1,x_2,x_3)&=&\Psi^*(x_1,x_2,x_3)~,\nn
\eea
and decompose the Fock space spanned by the functions (\ref{bel})
into the irreducible representations of the 
permutation group $S_3$ and time reflection $T$.
For given $(n_1,n_2,n_3)$ we define
\bea
\Psi_{nnn}^{(0)+}:=\Psi_{n n n}~,\nn
\eea
if all three indices are equal (type I), the three states $(m=-1,0,1)$
\bea
\Psi^{(m)+}_{nns}:=
{1\over\sqrt{3}}\sum_{k=0}^2 e^{-2km\pi i/3}
\left(P_{123}\right)^k\Psi_{n n s}~,\nn
\eea
when two indices are equal (type II), and the two sets of three states
$(m=-1,0,1)$
\bea
&&\Psi^{(m)\pm}_{n_1n_2n_3}:= \nn\\
&&\ \ \ \ {1\over\sqrt{6}}\sum_{k=0}^2 e^{-2km\pi i/3}\left(P_{123}\right)^k
 \left(1\pm P_{12}\right)\Psi_{n_1 n_2 n_3}~,\nn
\eea
if all $(n_1,n_2,n_3)$ are different (type III).
In this new orthonormal set of irreducible basis states 
$\Psi^{(m)\alpha}_{\bf N}$, 
the Fock representation of the Hamiltonian $H_0$ reads
\bea
\label{H_0irr}
H_0=\sum |\Psi^{(m)\alpha}_{\bf M}>
<\Psi^{(m)\alpha}_{{\bf M}}|H_0|\Psi^{(m)\alpha}_{\bf N}>
<\Psi^{(m)\alpha}_{\bf N}|~.\nn
\eea
The basis states $\Psi^{(m)\alpha}_{\bf N}$ are eigenfunctions of
$P_{123}$ and $P_{12}T$ 
\bea
P_{123}\Psi^{(m)\pm}_{\bf N}&=&e^{2m\pi i/3}\Psi^{(m)\pm}_{\bf N}~,\nn\\ 
P_{12}T\Psi^{(m)\pm}_{\bf N}&=&\pm \Psi^{(m)\pm}_{\bf N}~.
\eea
Under $P_{12}$ and $T$ separately, however, they transform into each other
\bea
P_{12}\Psi^{(m)\pm}_{\bf N}&=&\pm \Psi^{(-m)\pm}_{\bf N}~,\nn\\
T\Psi^{(m)\pm}_{\bf N}&=& \Psi^{(-m)\pm}_{\bf N}~.\nn
\eea
We therefore have the following irreducible representations. 
The singlet states $\Psi^{(0)+}$, the ``axial'' singlet states $\Psi^{(0)-}$, 
the doublets $(\Psi^{(+1)+};\Psi^{(-1)+})$ and the ``axial'' doublets 
$(\Psi^{(+1)-};\Psi^{(-1)-})$. 
Since the partner states of the doublets transform into each 
other under the symmetry operations $P_{12}$ or $T$, the corresponding
values of the energy functional are equal.

Due to this decomposition of the Fock space into the irreducible
sectors, the variational approach allows us to give
upper bounds for states in each sector.  
The values of the energy functional for the states in each irreducible
sector with the smallest number of knots  
${\cal E}[\Psi_{000}^{(0)+}]= 4.8676~ g^{2/3}$,
${\cal E}[\Psi_{100}^{(\pm 1)+}]= 7.1915~  g^{2/3}$,
${\cal E}[\Psi_{012}^{(0)-}]= 13.8817~ g^{2/3}$, and
${\cal E}[\Psi_{012}^{(\pm 1)-}]= 15.6845~ g^{2/3}$
give first upper bounds for the lowest energy eigenvalues 
of the singlet, the doublet, the axial singlet, and the 
axial doublet states.

In order to improve the upper bounds for each irreducible sector, 
we truncate the Fock space at a certain number of knots 
of the wave functions and search for the corresponding states
in the truncated space with the lowest value of the energy functional.
We achieve this by diagonalizing 
the corresponding truncated Hamiltonian $H_{\rm trunk}$ to
find its eigenvalues and eigenstates. Due to the orthogonality of the
truncated space to the remaining part of Fock space the value of the energy
functional (\ref{energyf}) for the  eigenvectors of $H_{\rm trunk}$
coincides with the $H_{\rm trunk}$ eigenvalues.  

Including all states in the singlet sector with up to $5$ knots 
we find rapid convergence to the following energy expectation values for 
the three lowest states 
$S_1,S_2,S_3$ 
\bea
{\cal E}[S_1] &=& 4.8067~ g^{2/3}\ (4.8070~ g^{2/3}) ,\nonumber\\
{\cal E}[S_2] &=& 8.2515~ g^{2/3}\ (8.2639~ g^{2/3}) ,\nonumber\\
{\cal E}[S_3] &=& 9.5735~ g^{2/3}\ (9.6298~ g^{2/3}) ,
\eea
where the number in brackets show the corresponding result for $4$ knots.
The lowest state $S_1$ is very close to the state $\Psi_{000}^{(0)+}$:
\bea
S_1 &=& 0.994562~ \Psi_{000}^{(0)+} + 0.0252533~ \Psi_{001}^{(0)+}\nonumber\\
    && 0.0216617~ \Psi_{002}^{(0)+} - 0.0970056~ \Psi_{110}^{(0)+}\nonumber\\
    && 0.0145832~ \Psi_{111}^{(0)+}. 
\eea
Similarly including all states in the doublet sector with up to $6(5)$ 
knots the
following energy expectation values for the three lowest states 
$D_1^{(\pm 1)},D_2^{(\pm 1)},D_3^{(\pm 1)}$
\bea
{\cal E}[D_1^{(\pm 1)}] &=& 7.1682~ g^{2/3}\ (7.1689~ g^{2/3}) ,\nonumber\\
{\cal E}[D_2^{(\pm 1)}] &=& 9.6171~ g^{2/3}\ (9.6394~ g^{2/3}) ,\nonumber\\
{\cal E}[D_3^{(\pm 1)}] &=& 10.9903~ g^{2/3}\ (10.9951~ g^{2/3}).
\eea
have been obtained.
Including all states in the axial singlet sector with up to $8(7)$ knots 
we find the following energy expectation values for the three lowest 
states $A_1,A_2,A_3$
\bea
{\cal E}[A_1] &=& 13.2235~ g^{2/3}\ (13.2275~ g^{2/3}),\nonumber\\
{\cal E}[A_2] &=& 16.6652~ g^{2/3}\ (16.7333~ g^{2/3}),\nonumber\\
{\cal E}[A_3] &=& 19.1470~ g^{2/3}\ (19.3028~ g^{2/3}).
\eea
Finally taking into account all states in the axial doublet sector with up 
to $8(7)$ knots 
we find the following energy expectation values for the three lowest states 
$C_1^{(\pm 1)},C_2^{(\pm 1)},C_3^{(\pm 1)}$
\bea
{\cal E}[C_1^{(\pm 1)}] &=& 14.8768~ g^{2/3}\ (14.8796~ g^{2/3}),\nonumber\\
{\cal E}[C_2^{(\pm 1)}] &=& 17.6648~ g^{2/3}\ (17.6839~ g^{2/3}),\nonumber\\
{\cal E}[C_3^{(\pm 1)}] &=& 19.9019~ g^{2/3}\ (19.9914~ g^{2/3}),
\eea
We therefore obtain rather good estimates for the energies of the lowest
states in the spin-0 sector. Extending to higher and higher numbers of
knots in each sector we should be able to obtain the low energy spectrum
in the spin-zero sector to high numerical accuracy.

In summary comparing our results for the first few states in all sectors,
we find that the lowest state appears in the singlet sector with energy
$4.8067~ g^{2/3}$ with expected accuracy up to three digits after the dot.
For comparison we remark that due to our boundary condition (\ref{bc1})
all our spin-0 states correspond to the $0^-$ sector in the work of
\cite{Martin} where a different gauge invariant representation of Yang-Mills 
mechanics has been used. Their state of lowest energy in this sector 
is $9.52~ g^{2/3}$. Furthermore in \cite{Giga}, using an analogy of $SU(N)$
Yang-Mills quntum mechanics in the large $N$ limit to membrane theory,
obtain the energy values $6.4690~ g^{2/3}$ and $19.8253~ g^{2/3}$ for
the groundstate and the first excited state.
  
Finally we remark that an analogous calculation shows that also the
groundstate energy of the Hamiltonian 
$H^\prime$ in Eq. (\ref{HH_0}) is lower than the value $E^\prime_{\rm sep}$ 
of (\ref{sepest}) for the lowest separable solution.  
\bigskip

The authors thank G. R\"opke for stimulating discussions.
The work of A.M.K. was supported in part by the Russian Foundation for
Basic Research under grant No. 98-01-00101 and
H.-P. P.  acknowledges support by the Deutsche Forschungsgemeinschaft
under grant No. RO 905/11-3.
Our collaboration is supported by the Heisenberg-Landau program, grant
HL-99-09.
%%%%%%%%%%%%%%%%%%%%%

\end{document}